\documentclass[namedreferences]{solarphysics}
\usepackage[optionalrh,solaenum,natbib,optionalrh]{spr-sola-addons} 
\usepackage{epsfig}    
\usepackage{graphicx}  
\usepackage{color}     
\usepackage{url}       
\usepackage{amssymb}
\usepackage{type1cm} 

\usepackage{lineno}    
\usepackage{setspace} 
\usepackage{cancel}
\usepackage{float}
\usepackage[colorlinks=true,citecolor=blue,breaklinks=true]{hyperref}

\newcommand{\Ha}{\hbox{H$\alpha$}}
\newcommand{\Ca}{\hbox{Ca\,{\sc ii}}}

\newcommand{\kmps}{\hbox{~km\,s$^{\rm -1}$}}

\newcommand{\arcsec}{\mbox{$^{\prime\prime}$}}%
\newcommand{\arcdeg}{\mbox{$^\circ$}}%
\newcommand{\Haa}{\Ha($\delta\lambda$=0.35\AA)}%
\newcommand{\Caa}{\Ca($\delta\lambda$=0.15\AA)}%
\newcommand{\vrms}{\hbox{$v_{\rm rms}$}}

\newcommand{\etal}{{\it et al.}}

\newcommand{\cf}{{\it cf.}}
\newcommand{\eg}{{\it e.g.}}
\newcommand{\etc}{{\it etc}}

\newcommand{\aap}{    {\it Astron. Astrophys.}}
\newcommand{\apj}{    {\it Astrophys. J.}}
\newcommand{\apjl}{   {\it Astrophys. J.}}
\newcommand{\apss}{   {\it Astrophys. Spa. Sci.}}
\newcommand{\nat}{    {\it Nature}}
\newcommand{\pasj}{   {\it Pub. Astron. Soc. Japan}}
\newcommand{\solphys}{{\it Solar Phys.}}
\newcommand{\ssr}{    {\it Space Sci. Rev.}}
\begin{document}
\date{\today}

\begin{article}
\begin{opening}

\title{Chromospheric Sunspot Oscillations in \Ha~and \Ca~8542\AA}

\author{Ram Ajor~\surname{Maurya}\sep Jongchul~\surname{Chae}\sep Hyungmin~\surname{Park}\sep Heesu~\surname{Yang}\sep Donguk~\surname{Song}\sep Kyuhyoun~\surname{Cho}}

\runningauthor{R. A. Maurya, \etal}
\runningtitle{Chromospheric Sunspot Oscillations in \Ha~and \Ca~8542\AA}

 \institute{Astronomy Program, Department of Physics and Astronomy, Seoul National University, Seoul 151-747, Republic of Korea.
               e-mail: \url{ramajor@astro.snu.ac.kr}\\
         }

\begin{abstract}
We study chromospheric oscillations including umbral flashes and running penumbral waves in a sunspot of active region (AR) using scanning spectroscopy in \Ha~and \Ca~8542\AA\, with the Fast Imaging Solar Spectrograph (FISS) at the 1.6 meter New Solar Telescope at Big Bear Solar Observatory. A bisector method is applied to spectral observations to construct chromospheric Doppler velocity maps. Temporal sequence analysis of these shows enhanced high-frequency oscillations inside the sunspot umbra in both lines. Their peak frequency gradually decreases outward from the umbra. The oscillation power is found to be associated with magnetic-field strength and inclination, with different relationships in different frequency bands. 
\end{abstract}


\end{opening}

\section{Introduction}
\label{S-Intro}

Umbral flashes (UFs) and running penumbral waves (RPWs) are long-known sunspot oscillation phenomena that have been studied extensively. The first is more three-minute, the latter more five-minute in character.  UFs were first discovered by \citet{Beckers1969} in \Ca~H and K filtergrams and spectrograms from sunspots. It has been suggested that UFs brightness forms in the local umbral gas during the compressional phase of a magneto-acoustic wave \citep[\eg,][]{Havnes1970, RouppevanderVoort2003} and recently it has been confirmed by \citet{Bard2010} from the NLTE simulations of \Ca~H line. RPWs were first studied by \citet{Zirin1972}. They appear as a sort of continuation of some of the flash waves. Main difference between the UFs and the RPWs is that the UFs are best seen in \Ca~H and K lines while RPWs are best seen in \Ha~Dopplergrams. Furthermore, UFs and RPWs have been interpreted as different manifestations of the same oscillatory phenomenon in a sunspot, combining upward-shock propagation with coherent wave spreading over the entire spot \citep{Zhugzhda1984a, RouppevanderVoort2003, Bloomfield2007}. The flashes are present only when the oscillation amplitude is sufficiently large ($\geq5$\kmps) but oscillatory motions are always present in nearly every umbra \citep{Moore1981}.

However, it is still unclear what determines the characteristic properties of the oscillations in sunspot chromospheres. How do the waves, that are associated with these oscillations, propagate in the magnetized atmosphere? What is the spatial variation of frequency across the sunspot and in features of different spatial scales? The oscillations associated with waves are important to study since they carry some information on the properties of the region from which they originate and through which they propagate. Furthermore, the rapidly evolving field of local helioseismology, to investigate the sub-surface structure and dynamics, uses the oscillation properties. 

UFs and RPWs appear more clearly in Dopplergrams than in intensities. In many earlier studies the chromospheric oscillations were analyzed by constructing Dopplergrams from filtergrams at fixed wavelengths in the red and blue wings. In this paper, simultaneous spectral observations in the \Ha~and \Ca~8542\AA~lines have been used to construct Doppler velocity maps (hereafter Doppler maps) from the bisectors of these two lines. The rest of the article is organized as follows: We describe the observational data in Section~\ref{S-Data} and present the methods of analysis in Section~\ref{S-Analys}. Results and discussions are given in Section~\ref{S-ResDisc}. Finally, Section~\ref{S-SumConc} is devoted to the summary and conclusions.  

\section{Observations}
\label{S-Data}

We observed AR NOAA 11242 on 30 June 2011. It was located at heliographic latitude 17\arcdeg\,N and longitude 29\arcdeg\,W, and consisted of a well-developed sunspot of southern polarity with mean magnetic field strength 1200\,G. The sunspot was surrounded by a number of small magnetic fragments of northern polarity. AR was well observed by the FISS instrument \citep{Chae2012}, and the Helioseismic and Magnetic Imager \citep[HMI: ][]{Schou2012} on board the {\it Solar Dynamics Observatory} (\,SDO\,).

FISS is a slit spectrograph taking spectra with rapid-scan capability. It observes the solar chromosphere simultaneously in two spectral bands centered around the lines \Ha~and \Ca~8542\AA~(hereafter \Ca) with spectral resolutions of 0.045\AA~and 0.064\AA, respectively. The pixel resolution at both wavelengths is $\approx$\,0.16\arcsec\,pixel$^{-1}$. 

The FISS data cube, with a field of view of 24\arcsec~of scan width (total number of scans 150) and 41\arcsec~of slit length, covered the entire sunspot of AR NOAA 11242. The observations were taken relatively under good seeing conditions during 17:51:38\,--\,18:55:42\,UT, except for a gap during 18:04:45 -- 18:18:10\,UT, with the scan step sampling, timing, and cadence of 27\,s, 130\,ms and 30\,s, respectively, in both  spectral bands.

In order to study the association between chromospheric oscillations and magnetic fields, we have used the HMI vector-magnetograms: field strength, azimuth angle, and angle of inclination. The angle of inclination is measured from the line-of-sight and azimuth from $+Y$ direction of the heliocentric coordinates (X,Y) with from apparent disk center through the solar poles.

\section{Data Reduction}
\label{S-Analys}

Out of 105 cubes for \Ha~and \Ca~each, seven were discarded being of low quality from failure of adoptive optics locking. The data are processed in two main steps: pre-processing, and post-processing. In the first step, we corrected the data for  bias, dark, flat and slit patterns. In the second step the wavelength scales were calibrated. Finally the processed data are compressed using principal component analysis \citep[PCA:][]{Pearson1901} which is useful specially when spectral profiles are similar to one another \citep{Rees2000}. PCA compression suppresses the noise to a substantial amount without much loss of information. More detail on the processing and PCA compression can be found in \citet{Chae2012}. 

\begin{figure} \centering
\includegraphics[width=1.0\textwidth,clip=,bb=38 22 479 473]{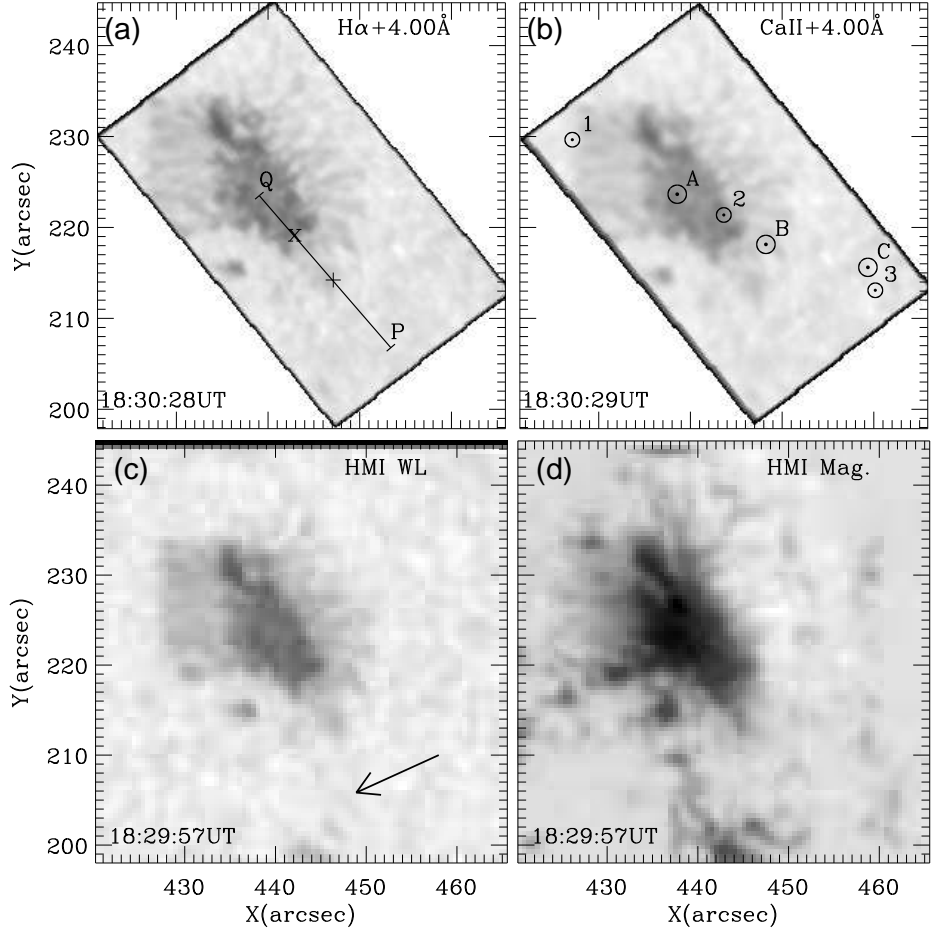}
\caption{Image mosaic of the AR NOAA 11242 on 30 June 2012 in different wavelengths. The upper panels show the raster images at wavelengths 4.0\AA~from  \Ha~(a) and \Ca~(b) line center. The line PQ (a) marks reference locations for the analysis. The locations marked by the symbols ``$\times$'' and ``+'' represent the approximate boundaries of the umbra and the penumbra, respectively. Small circles labeled with integer numbers (b) mark the locations for which time information is shown in Figure~\ref{F-DelCad} while big circles labeled with A, B and C represent three locations as in Figure~\ref{F-DopMap}. The lower panels show the HMI white-light (c) and magnetogram (d) images observed near the time of the FISS observation (upper panels). The arrow in panel (c) denotes the direction of the solar disk center.} \label{F-ImMos}
\end{figure}

\subsection{Alignment}
\label{Sb-DataAlign}

Image rotation at the Coud\'e focus, telescope guiding errors, errors in slit positioning, and seeing cause shifts between successive scans. We used the far-wing parts of the scans as references. The successive scans were first derotated and then aligned by cross-correlation to the reference scan, and the \Ha~and \Ca~scans were also aligned by cross-correlation. The error due to slit - positioning is found to be $\pm0.16$\arcsec. Typical values of the net shifts in scan $x$-and $y$-directions were found to be 8 pixel and 2 pixels, respectively. The FISS data were aligned with the HMI images, by manual feature and pattern matching, into solar ($X$, $Y$) coordinates. 
    
Figure~\ref{F-ImMos} shows a mosaic of the aligned FISS and HMI images of AR NOAA 11242 on 30 June 2011. The FISS \Ha~(a) and \Ca~(b) images are constructed from the spectral data at wavelength 4.0\AA~from line center. The bottom panel shows the HMI white light (c) and magnetogram (d) images of the same region observed at nearly the same time as the FISS images (specified in each panel bottom). The line PQ drawn in panel (a) marks the reference locations along which the characteristics of oscillations have been analyzed (Section~\ref{S-ResDisc}). The arrow in the bottom-left panel shows the direction of solar disk center. 

\begin{figure} \centering
\includegraphics[width=1.0\textwidth, clip=,bb=19 11 375 375]{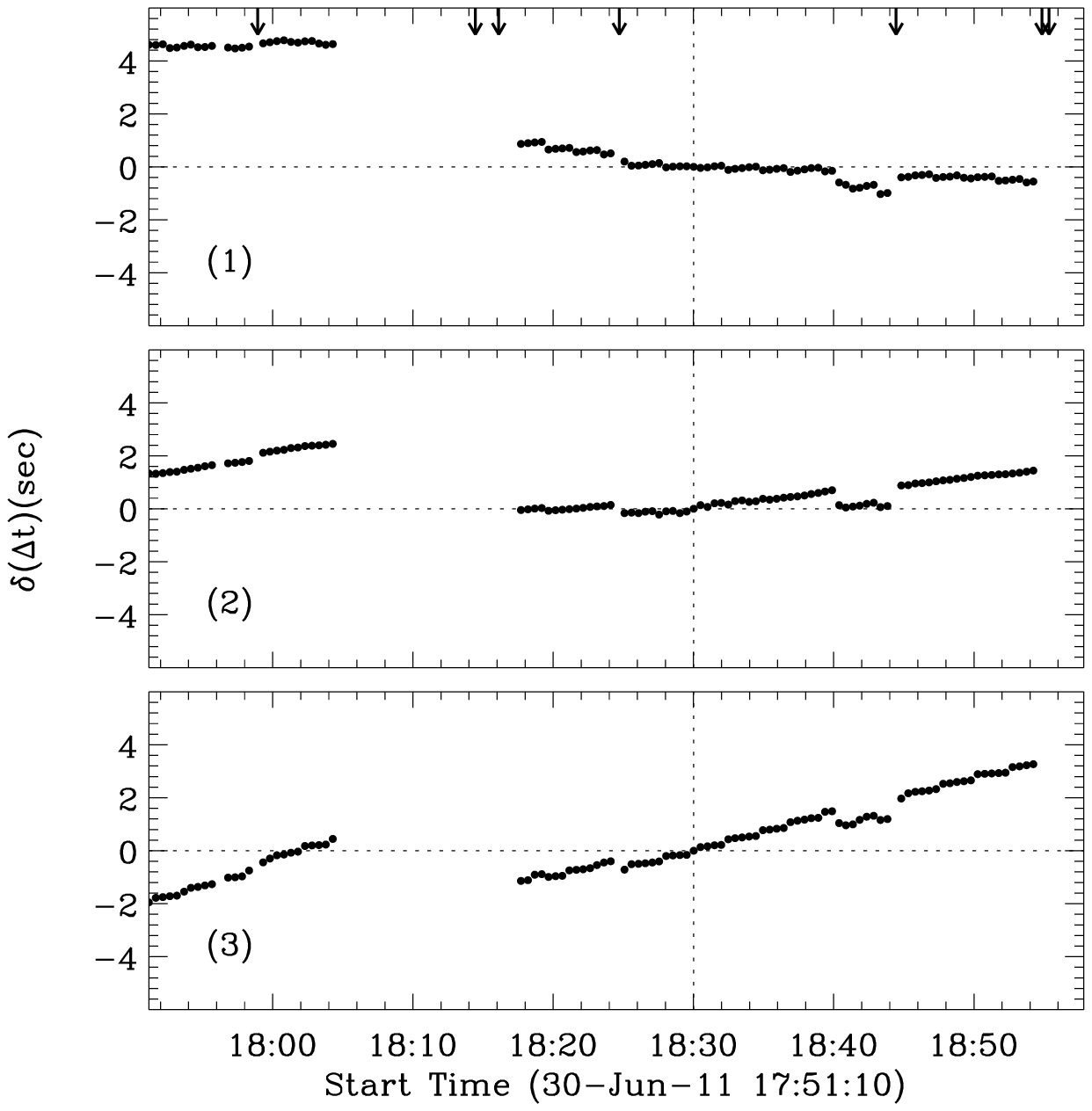}
\caption{Timing shifts due to image rotation and scanner shifts at three locations marked in Figure~\ref{F-ImMos}(b). Vertical dotted line specifies the reference image. Arrows in the upper panel mark the locations of bad-AO scans.} \label{F-DelCad}
\end{figure}

\subsection{Timing Correction}
\label{Sb-TimCad}

Both the image rotation and the image shifts between successive scans that were corrected by derotation and alignment imply that the timing per pixel has no longer a fixed cadence. We determined the actual observing times per pixel from the alignment vectors. 

Figure~\ref{F-DelCad} shows the temporal variation in cadence, $\delta(\Delta t)$, for 3 locations (1, 2 and 3 marked with circles in Figure~\ref{F-ImMos}b). There is a data gap between 18:04:45 -- 18:18:10\,UT, also the images have large shift and rotation. In addition,  $\delta(\Delta t)$ varies in the range ~$\pm5$\,s, for the total observation period of 63 minutes. The deviation $\delta(\Delta t)$ for the point 1 (3) becomes more negative (positive) with time which suggests that the pixel corresponding to this position was observed earlier (latter).  Note that we have only used the data sets starting from the time 18:18:10\,UT, for the oscillation studies. Also the 7 bad-AO scans were taken out.

\subsection{Bisector Measurement, Doppler Maps and Power Maps}
\label{Sb-TempEvol}

We compute the Doppler velocity using the bisectors \citep{Gray1976, Dravins1981} of spectral line profiles.  The locus of bisectors represents the asymmetry of the spectral line around the line center. For the bisector for a given spatial location, one may sample the profile asymmetry at a given intensity level \citep{Cavallini1986, Keil1981}, at a given separation from the nominal line center wavelength or the per-pixel measured minimum wavelength \citep{Bhatnagar1972a, vonUexkuell1983, Tziotziou2002, Tziotziou2007}, and lambdameter measurement at given chord length  \citep{Slaughter1972, Stebbins1987}. We prefer to measure Doppler velocities with the lambdameter method. 

Of course lambdameter method has problem, along with the first two methods, in the case when the Doppler width is changing, \eg, spectral lines are wider in the sunspot umbra than quiet region. There is a possibility of width variation with time for a given location due to energetic activities. However, it is difficult to remove the width related variation in the Doppler velocity for every pixel in the time series. To minimize this effect, we have chosen an optimum value of the bisector chord,  $\delta\lambda=$ 0.35\AA(0.15\AA) for \Ha(\Ca). These chords were applied to all the spatial pixels to compute the Doppler maps (see Figure~\ref{F-DopMap}, bottom row).  To average over remaining image distortions due to seeing, guiding errors, \etc., we applied spatial $3\times3$ boxcar smoothing. For the power analysis of Doppler maps, we prefer not to interpolate the data to equal temporal sampling but instead used the Lomb-Scargle periodogram (LSP) technique \citep{Lomb1976, Scargle1982}.

\begin{figure}[ht] \centering
\includegraphics[width=1.0\textwidth, clip=,bb=58 15 505 234]{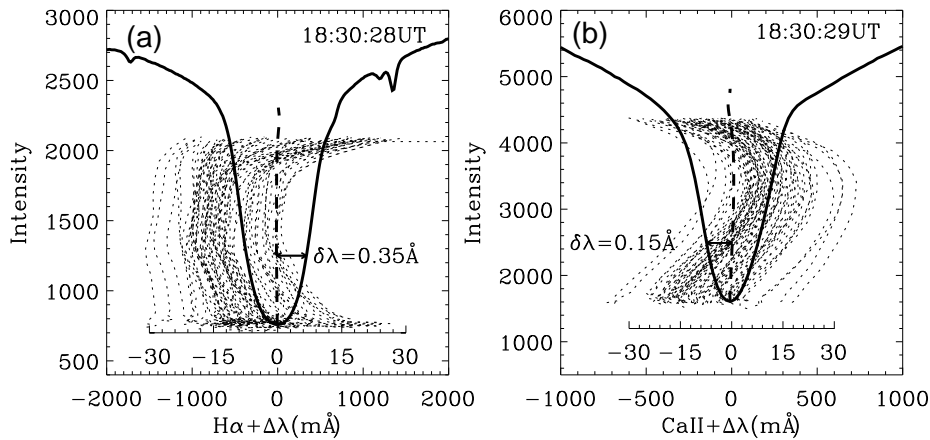}
\caption{Time-average intensity profiles (thick-solid) and bisectors (thick-dashed) for the location B (see Figure~\ref{F-DopMap}). Thin-dashed curves, plotted with respect to $x$-axes drawn near lines' cores, represent  bisectors on a magnified scale (50 times for \Ha~and 20 times for \Ca) sampled at the same location at different times. The horizontal arrow marks bisector chord $\delta\lambda=0.35$\AA($0.15$\AA)~for \Ha(\Ca).} \label{F-bisect}
\end{figure}

\section{Results and Discussions}
\label{S-ResDisc}

\begin{figure}[ht]\centering
\includegraphics[width=1.0\textwidth, clip=,bb=38 19 521 480]{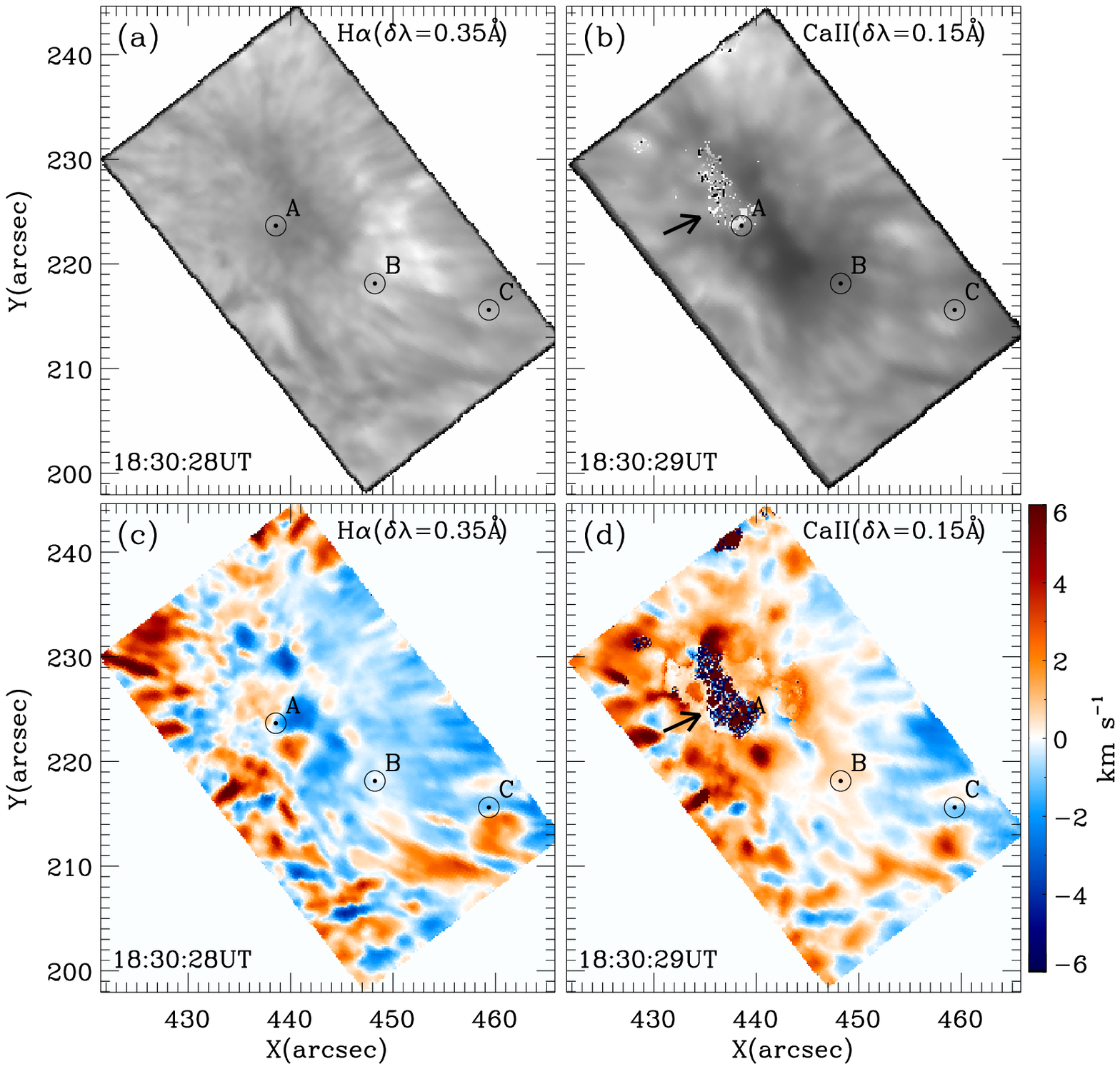}
\caption{Simultaneous intensity (upper row) and Doppler velocity (lower row) maps of AR NOAA 11242 computed using bisector method for the \Haa~and \Caa.  Arrows mark the umbral locations where Doppler velocity measurement in \Ca~failed due to emission features in umbral region. The positions A, B and C labeled with circles represent the locations for which wavelength-time maps are shown in Figure~\ref{F-WvTimMap}.}\label{F-DopMap}
\end{figure}

Figure~\ref{F-bisect} shows time-averaged intensity profiles (thick-solid) and bisectors (thick-dashed) of \Ha~(a) and \Ca~(b) lines for location B (see Figure~\ref{F-ImMos}(b) and Figure~\ref{F-DopMap}). Thin-dashed curves show the magnified bisectors (50 and 20 times respectively for the \Ha~and \Ca) at different times for the same location B, and can be read from the $x$-axes drawn near the line cores. 

The bisectors for the \Ha~line is ``C'' shaped while for the \Ca~line they are inverse of ``C'', and represent blue and red asymmetry, respectively, in line profiles. \citet{Uitenbroek2006} has also reported inverse ``C'' shaped bisectors and a strong red asymmetry in the \Ca~line. The amount of curvature or equivalently the velocity covered by the bisector is a measure of asymmetric flows. 

Figure~\ref{F-DopMap} (bottom row) shows samples of chromospheric Doppler maps  of the \Ha~and \Ca~lines respectively constructed  at bisector chords $\delta\lambda=0.35$\AA~(c) and $\delta\lambda=0.15$\AA~(d), where the upper panels are  intensity maps obtained from the mean of the intensities of the two wings at the chord intersections. The Doppler velocities in these maps range $\pm$6\,km\,s$^{-1}$, where negative/positive velocity shows  upward/downward plasma motions. Patterns associated with running penumbral waves, in \Ha~and \Ca, can be seen in the Doppler maps. The overall red-blue asymmetry of the Doppler maps is caused by off-disk-center viewing. The blue in the limbward part of the penumbra indicates the line-of-sight component of the reverse Evershed flow. It is less evident on the center side, although filaments there would have better alignment because this spot has no penumbra in the centerside quadrant.

\begin{figure}[h] \centering
\includegraphics[width=1.0\textwidth, clip=,bb=54 18 493 239]{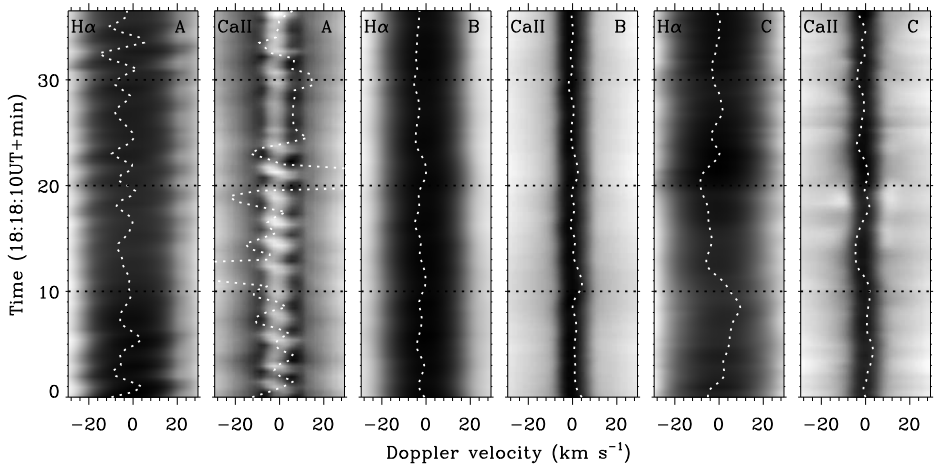}
\caption{Intensity as a function of wavelength and time for the \Ha~and \Ca~lines at three spatial locations, A, B and C, respectively for the umbra, penumbra and super-penumbra (see Figure~\ref{F-DopMap}). Dotted curves, passing through the line's centers, show the Doppler velocities (magnified by factor of 5 for all the panels except the panel 2 from the left) at corresponding locations.} \label{F-WvTimMap}
\end{figure}

Near the umbra in \Ca, there are pixels with high velocities (mark by arrow) which are artifacts and occur due to failure of the method adopted for the Doppler velocity determination. Similar kinds of Doppler velocity and magnetic field artifacts have also been reported earlier in the measurements obtained from the Michelson Doppler Imager \citep{Maurya2009, Maurya2010} and the HMI \citep{Maurya2012}. We found that these pixels are associated with emission features of UFs marked by arrows in the intensity images. This problem was already described by \citet{Kneer1981}, and also by \citet{Tziotziou2006}. UF emission features in \Ca~can also be seen in Figure~\ref{F-WvTimMap}.  

Figure~\ref{F-WvTimMap} shows the intensity variations in \Ha~(panels, 1, 3 and 5 from the left) and \Ca~(panels, 2, 4 and 6 from the left) as a function of time and wavelength for the three spatial locations A, B and C corresponding to the umbral, penumbral and outer-penumbral regions, respectively. The Doppler velocities for bisector chords lengths of 0.35\AA(\Ha) and 0.15\AA(\Ca) for corresponding positions are over plotted (dotted curves). Here note that the computed Doppler velocities are much smaller than the spectral ranges as shown in the the abscissa axes, and therefore we magnified their values by factors of 5 for both the \Ha~and \Ca, respectively, except the panel 2 from the left.

For the position A, the \Ca~line core shows emission features of UFs and the Doppler velocities computed through bisectors may be inaccurate. Nevertheless, the Doppler velocities for all locations vary quasi-sinusoidally with time which represents the oscillatory motions of the chromospheric plasma emitting the \Ha~and \Ca~lines. There is an interesting pattern of different periods when we move from the panel A to C, corresponding to different locations. The period of oscillations increases from A to C. 

\begin{figure}[ht] \centering
\includegraphics[width=1.00\textwidth, clip=,bb=35 15 565 494]{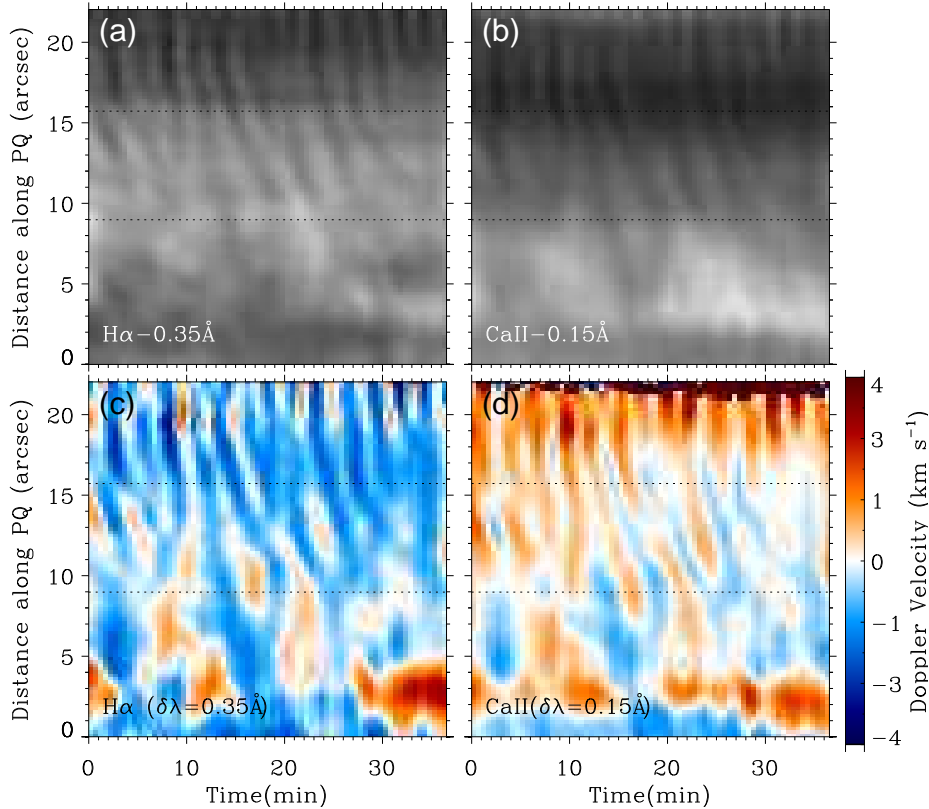}
\caption{Space-time maps of intensity (upper row) and Doppler velocity (lower row), in \Ha~(left column) and \Ca~(right column), along the reference line PQ (see Figure~\ref{F-ImMos}a). Dotted horizontal lines around at 16\arcsec~and 9\arcsec~represent the approximate boundaries of the umbra and the penumbra, respectively.} \label{F-RunPenWv}
\end{figure}

Figure~\ref{F-RunPenWv} shows space-time intensity maps (upper row) for  \Ha-0.35\AA~(a) and \Ca-0.15\AA~(b), and Doppler maps (lower row) for the \Haa~(c) and \Caa~(d) along the line PQ (see Figure~\ref{F-ImMos}a). It is evident that the contrast of the intensity and Doppler velocity variations in the umbral-penumbral region is large. Also the contrast in \Ha~is larger than in \Ca. In the umbral region, the ridges are almost vertical and show sinusoidal patterns of intensity and Doppler variations with time. Near the umbral boundary, there is a sharp change in the slope of the ridges. The intensity and Doppler velocity contrasts in the ridges decay with distance from the umbral boundary to outer edge of the penumbra, and it is hard to distinguish intensity ridges further. This shows that the RPWs decay in the super-penumbral regions of the sunspot. 

From the raster images, we found that the amplitude of RPWs gradually decreases from the line centers. It is difficult to identify the RPWs beyond the 0.75\AA~in the \Ha~and 0.50\AA~in the \Ca. Also, the amplitude of RPWs decreases from the umbral boundary to outer edge of the penumbra which is consistent with the earlier reports \citep{Giovanelli1972,Zirin1972}. 

\begin{figure}[ht] \centering
\includegraphics[width=1.0\textwidth, clip=,bb=41 62 524 445]{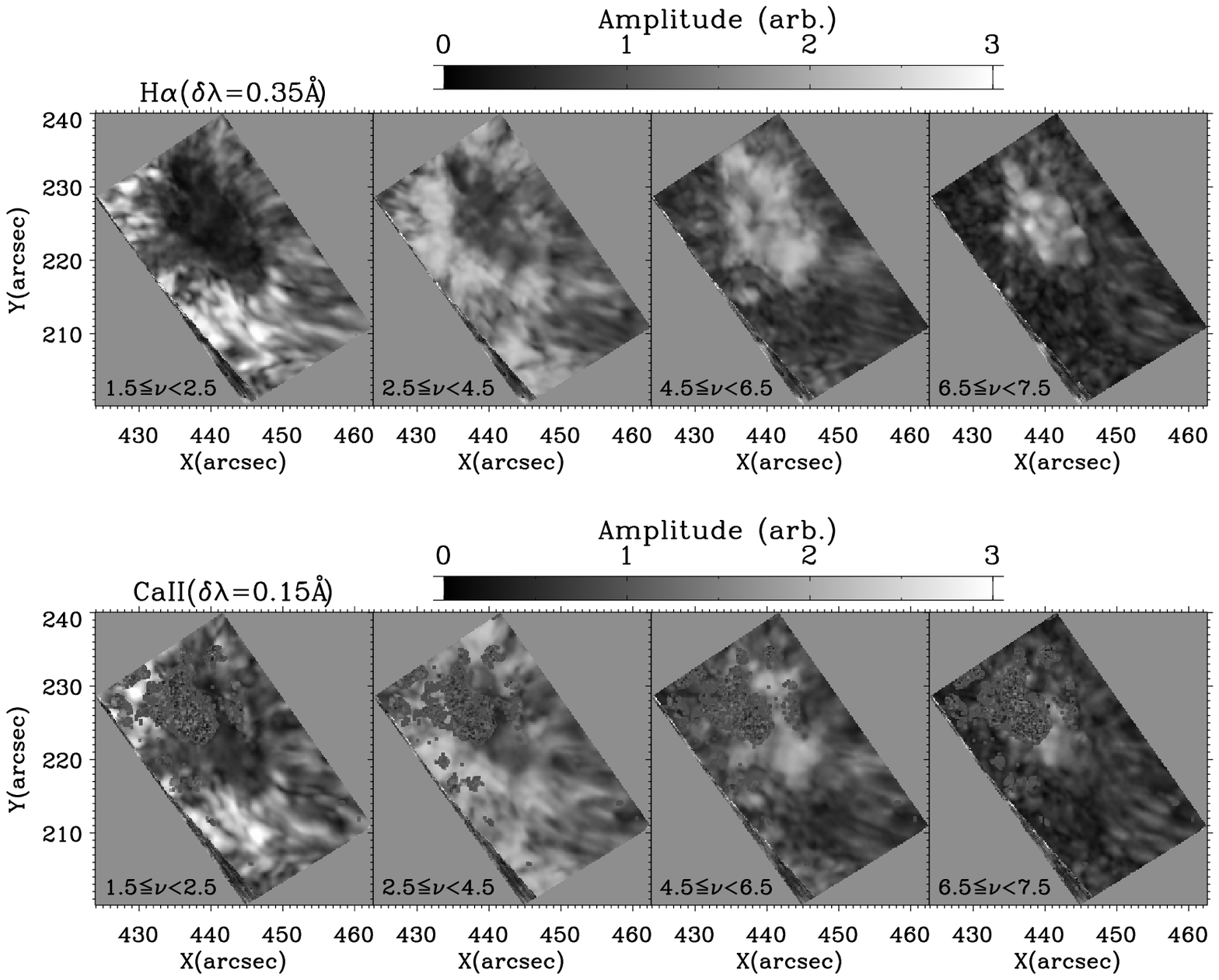}
\caption{Average oscillation power maps in different frequency bands in mHz (from the left to the right) for the \Haa~(upper) and \Caa~(lower).}\label{F-PwMapFrq}
\end{figure}

The space-time Doppler panels of Figure~\ref{F-RunPenWv} show that the RPWs have much more blue than red in \Ha, the reverse in \Ca~but no such asymmetry  appears in \citet{Tziotziou2006, Tziotziou2007}. However, their traces also have larger amplitude for \Ca~in the umbra plus a shift between largely negative for \Ha~and largely positive for \Ca.  
 
Figure~\ref{F-PwMapFrq} shows a sample of oscillation power maps, for the H$\alpha$($\delta\lambda$=0.35\AA) (upper) and \Caa (lower), averaged over different frequency bands. For comparison, all the maps are shown in the same amplitude range (see color scales in the top). It is evident that there is a large power in the sunspot umbra at higher frequency band ($4.5\leq\nu<7.5$\,mHz) than the lower frequency band ($1.5\leq\nu<4.5$\,mHz) while in the outer penumbra the power is large at lower frequencies than in higher frequencies. This is consistent with earlier reports, \eg, \citet{RouppevanderVoort2003}. This pattern exists in both the spectral bands. The high frequency oscillations in \Ca~are found to be concentrated in smaller regions than in \Ha, although there are some bad data points in the umbra of \Ca~maps due to artifacts in the Doppler velocity images (\cf~Figure~\ref{F-DopMap}). 

\begin{figure}[ht]\centering
\includegraphics[width=1.0\textwidth, clip=,bb=108 21 501 398]{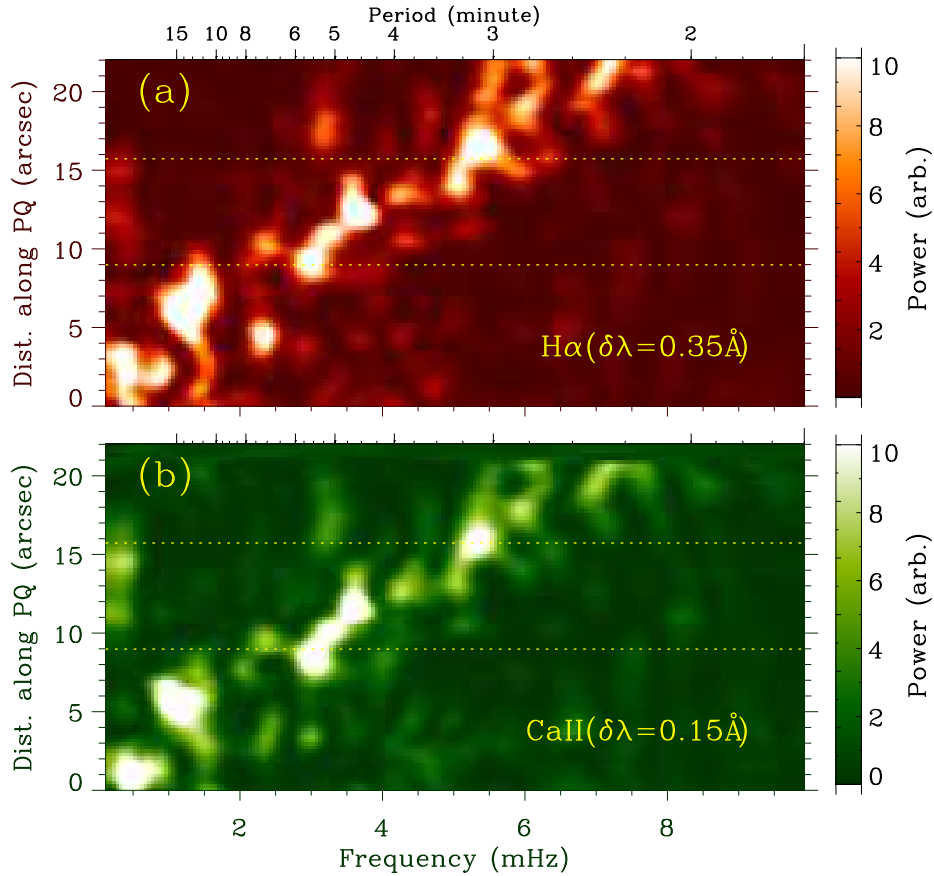}
\caption{Oscillation power maps along the line PQ (see Figure~\ref{F-ImMos}a) for the Doppler velocities of \Haa~and \Caa. Dotted lines mark the boundaries as shown in Figure~\ref{F-RunPenWv}.} \label{F-SpacLinePw}
\end{figure}

Figure~\ref{F-SpacLinePw} shows the oscillation power as a function of space (along the line PQ in Figure~\ref{F-ImMos}(a)) and frequency. There is a clear linear trend in the maximum power variation from the umbra to outward. The oscillations in the three and five minutes bands along with the features at other frequencies can be seen in both maps. 

\begin{figure}[ht] \centering
\includegraphics[width=0.786\textwidth, clip=,bb=46 28 321 458]{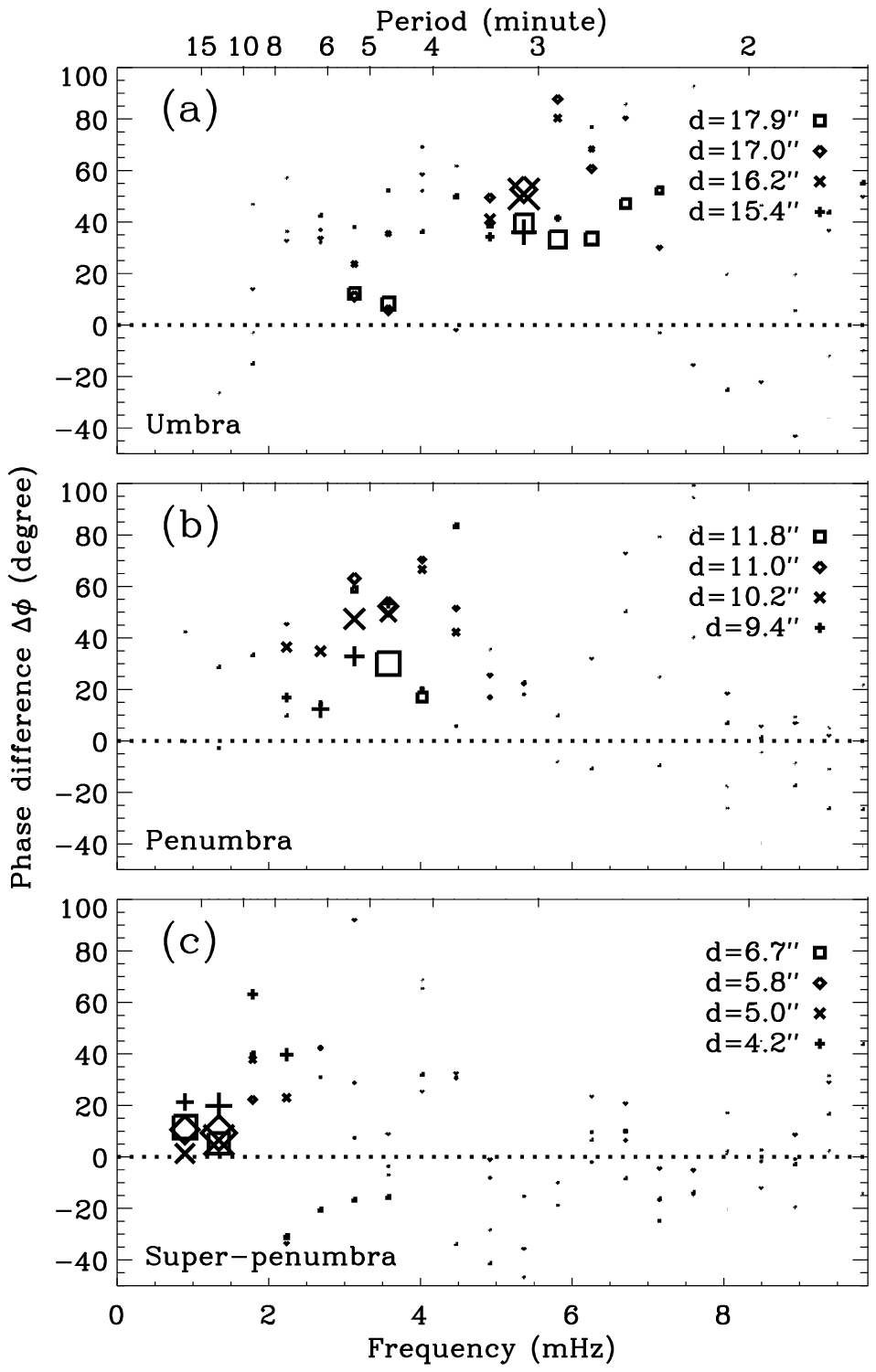}
\caption{Phase difference ($\Delta\phi$) in \Caa~and \Haa~Doppler velocities of: (a) umbra, (b) penumbra, and (c) superpenumbra, at four locations marked by symbols, $\Square$, $\Diamond$, $\times$, and + along the reference line PQ (Figure~\ref{F-Vrms}a), where \Caa~is taken as leading. The distance $d$ is measured from the point P along PQ. The symbol sizes represent the cross-spectral power between the two Doppler time series.} \label{F-PhaseDiff}
\end{figure}

In order to study the nature of the waves in the sunspot, we performed the Fourier phase difference analysis between the two Doppler signals obtained for \Ha~and \Ca. The Dopplershift measurements along PQ were interpolated to fixed-interval (30\,s) sampling for this purpose.

\begin{figure}[ht] \centering
\includegraphics[width=1.0\textwidth, clip=,bb=67 42 462 392]{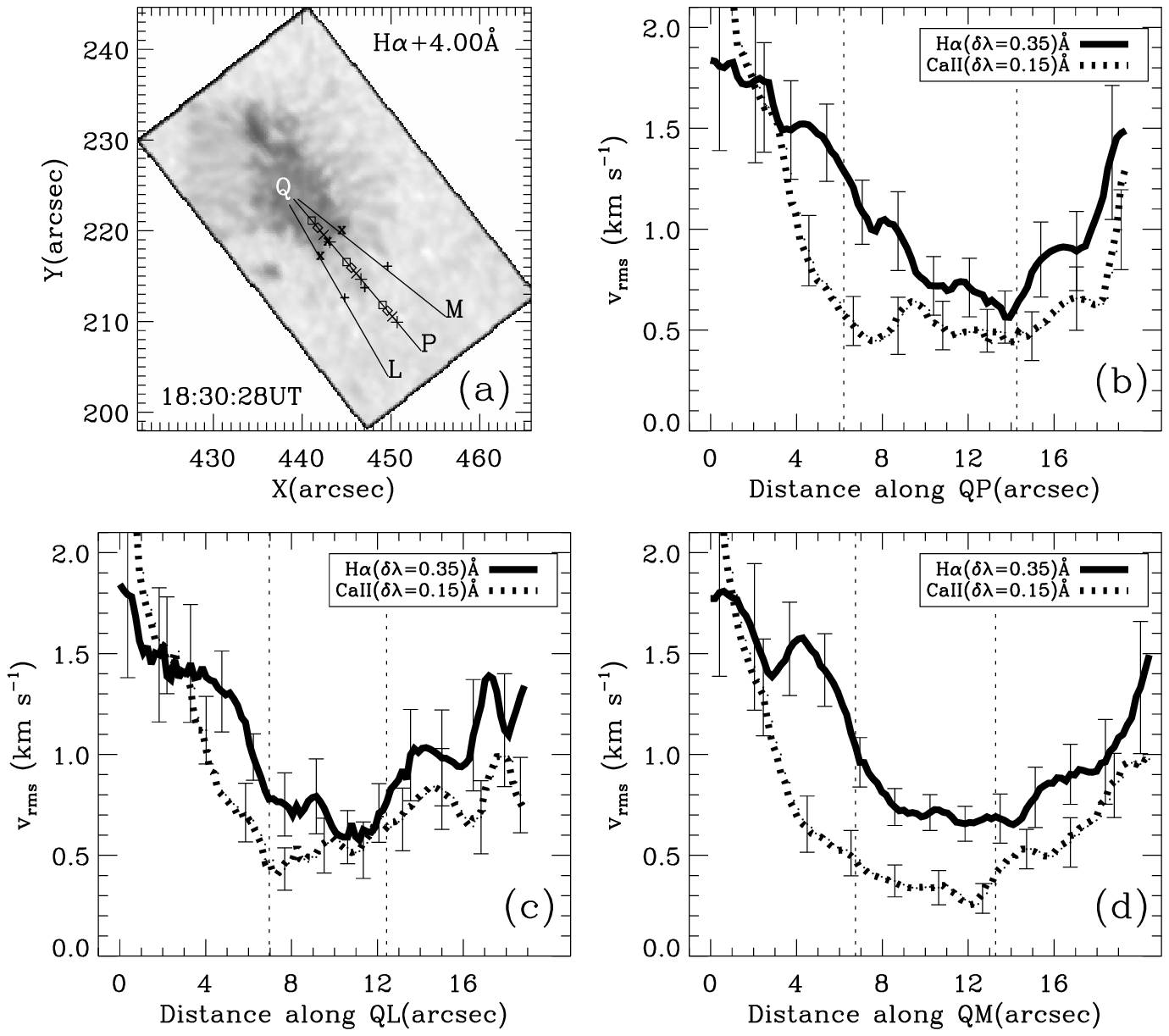}
\caption{Variation in \vrms~velocities of \Haa~and \Caa~along the reference lines QN(b), QM(d) and QL(c) marked in the panel (a). Vertical dotted lines show the approximate umbral and penumbral boundaries marked by symbols ``$\times$'' and ``+'', respectively, over the solid curves in the panel (a). Other symbols ``$\Square\Diamond\times$+'' in the umbra, penumbra and superpenumbra of the sunspot mark the locations along the line PQ (same as in Figure~\ref{F-ImMos}a) for phase analysis in Figure~\ref{F-PhaseDiff}.}\label{F-Vrms}
\end{figure}

Figure~\ref{F-PhaseDiff} shows the phase difference between the \Haa~and Ca{\sc ii} ($\delta\lambda$=0.15\AA) Doppler velocities as function of frequency for four locations in the umbra(a), penumbra(b) and superpenumbra (c); locations are marked in the panel (a) of Figure~\ref{F-Vrms} by symbols, $\Square$, $\Diamond$, $\times$, and +. Symbol sizes represent the cross-spectral power between the two Doppler velocities. The average phase difference in the frequency bands 5 and 3\,mHz are around 40\arcdeg~where cross-spectral powers are significant. The phase difference shows that the UFs and RPWs are caused due to upward propagating MHD waves. The phase difference in the superpenumbral region is about 10\arcdeg~at lower frequency $\sim$1\,mHz where significant power exists. This difference may caused due noise and unequal sampling in the original data.

In order to ascertain the spatial variation in the oscillation power from umbra to outward, we have computed the root-mean-square (RMS) velocity (\vrms) along the line PQ and results are shown in Figure~\ref{F-Vrms}. Dotted vertical lines correspond to approximate umbral and penumbral boundaries marked by symbols ``$\times$'' and ``+'', respectively, in the panel (a) of Figure~\ref{F-Vrms}. It is evident that \vrms~is large in the umbra of both the \Haa~and \Caa~and decreases (increases) in the penumbra (superpenumbra). In the inner umbra, \Ca~\vrms~for point Q rises considerably above the 2.1\,km\,s$^{-1}$ axis cutoff which is caused by combined effects of high Doppler velocities, and large uncertainty in the Dopplershift measurements. In the outer umbra, \vrms~of \Ha~is larger than \vrms~of \Ca. This may be due to the difference in formation height between the two spectral lines or other difference in response. For example, \Ca~may have more opacity in cool post-shock gas than \Ha~(which gets large opacity only in hot gas), so that \Ca~feels cool post-shock downdrafts and \Ha~does not. \Ca~then may get the K2V-like shock grain pattern from shock interference along the line of sight. Towards P the LOS alignment is lost. The large $\vrms$ in umbra consists mostly of high-frequency power. The high-frequency components likely describe sawtooth shocks (see Figure~\ref{F-WvTimMap}).

\begin{figure}[ht] \centering
\includegraphics[width=0.99\textwidth, clip=,bb=60 28 370 474]{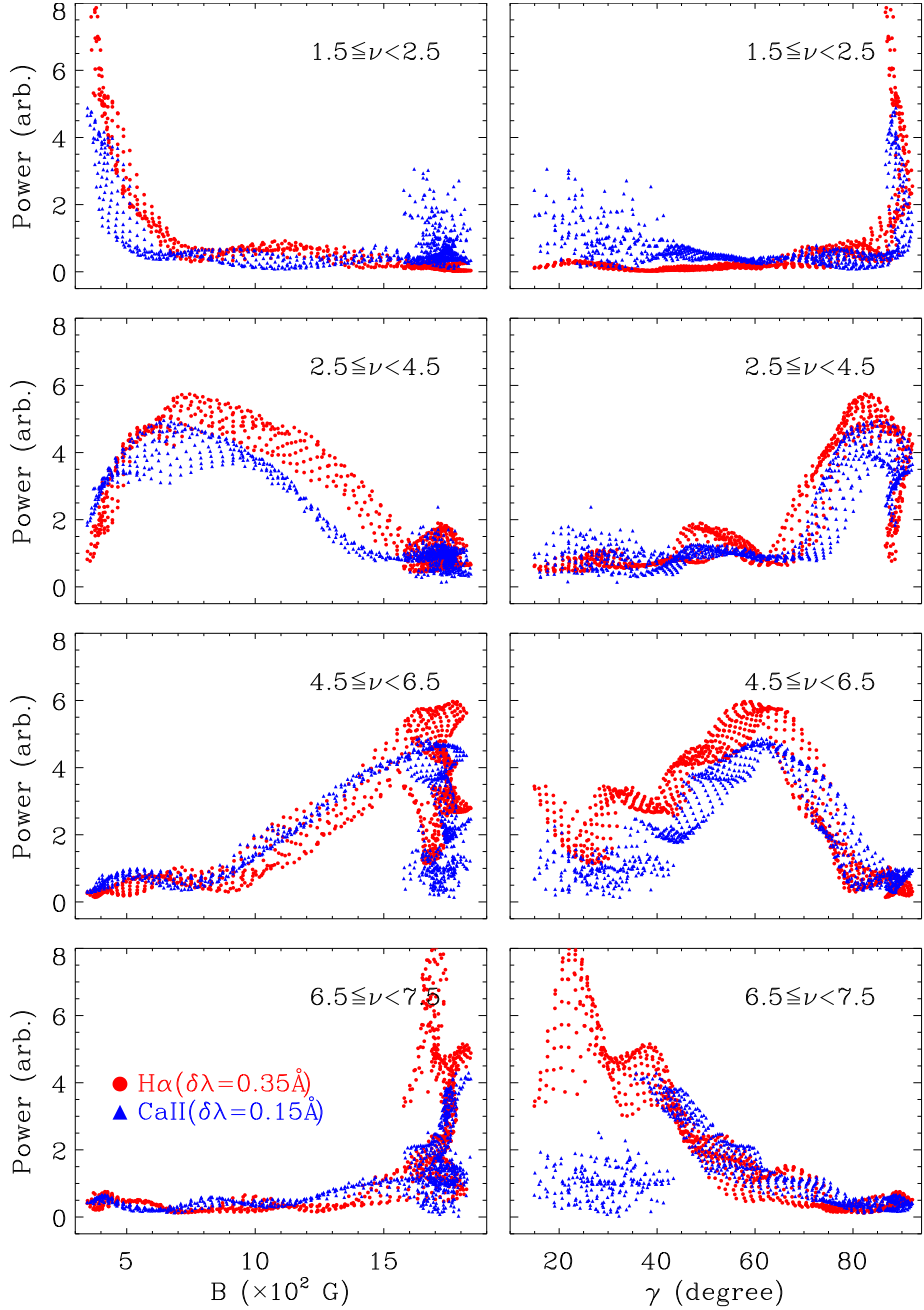}
\caption{Variation of the oscillation power in different frequency bands (from top to the bottom) with: (left) total magnetic field ($B$), and (right) inclination angle ($\gamma$) measured from a rectangular area along line PQ of width 10 pixels.} \label{F-MagPwDistrib}
\end{figure}

Figure~\ref{F-MagPwDistrib} (left column) shows the association between oscillation power and corresponding magnetic field strength $B$ from a rectangular area along line PQ of width 10 pixels. At lower frequencies ($1.5\leq\nu<2.5$\,mHz) the power is largest in areas with smaller $B$ which correspond to the regions of outward penumbra, and rapidly decreases with $B$. In the frequency band $2.5\leq\nu\leq4.5$\,mHz, the power initially increases up to $B\approx700$\,G and then decreases with $B$. In the frequency band $4.5\leq\nu\leq6.5$\,mHz the oscillation power increases with $B$ and becomes maximum around 1600\,G then decreases rapidly. In the frequency band $6.5\leq\nu\leq7.5$\,mHz most of the power is concentrated near the very high $B$ ($>1500$\,G) which correspond to the umbral region of the sunspot while the power is nearly constant for the $B$ range 300-1200\,G.

For the study of association between the oscillations and the field inclinations, we computed the inclination from the vertical to the surface using relation $\gamma=\cos^{-1}(B_{\rm r}/|\mathbf{B}|)$, where, $B_r$ is the radial component of the magnetic field $\mathbf{B}$. Uniform shear method is used \citep{Moon2003} to resolve the 180\arcdeg~ambiguity in the azimuth angle.

Figure~\ref{F-MagPwDistrib} shows the variation in the oscillation power with magnetic field inclination $\gamma$ corresponding to the left column of magnetic field strength $B$. It is evident that the power shows reverse trend with inclination angle than the field strength. In the lower frequency band ($1.5\leq\nu<2.5$\,mHz), most power is concentrated at higher $\gamma>80$\arcdeg. The peak of power distribution shifts toward lower inclination for higher frequency bands. For instance in the frequency band $6.5\leq\nu<7.5$\,mHz, most of the power is concentrated in the inclination region 10\arcdeg -- 30\arcdeg. 

The opposite relation of inclination and field strength with the power is obvious from the magnetic field distribution in the sunspot umbra; we have strong field with small inclination. From the umbra to outward the field strength increases while inclination decreases. Figure~\ref{F-MagPwDistrib} shows that the oscillation power is associated with the field strength and the inclination angle both but they behave differently in different frequency bands. 

\section{Summary and Conclusions}
\label{S-SumConc}

We studied the nature of running penumbral waves using \Ha~and \Ca~Doppler images constructed with a bisector method from area scans with the FISS spectrometer. We found that the RPWs are easily seen in the intensity images constructed near the core and their amplitude decay with wavelength from the line center to outward. Also their amplitude decreases from the umbra to outward. These results are consistent with earlier results about the RPWs \citep{Zirin1972, Christopoulou2000, Tziotziou2006}. The running penumbral waves decay near the outer boundary of penumbra.

We found that the chromosphere umbra of the sunspot shows large RMS velocity (\vrms) in both spectral lines. From the umbra to outward the \vrms~gradually decreases, and there is no distinct boundary between the \vrms~of the umbra and penumbra, however, \vrms~has minimum in the penumbral region. There is another interesting pattern seen in the \vrms~of \Ha~and \Ca~lines, \vrms~is smaller in the \Ca~than \Ha~away from the umbral center. These results reveal that the \Ha~and \Ca~are formed differently in the shocks that make up UFs. Furthermore \vrms~is related to the oscillation power (power $\propto$ \vrms$^2$). That is, we have larger power in the umbra than in penumbra, also, the power is larger in the \Ha~than in \Ca. 

Our times series analysis of chromospheric Doppler maps shows high - frequency power in the umbra of both the spectral lines, \Ha~and \Ca, which is also evident from the \vrms~distribution. This study shows that the total power gradually decreases with frequency from the umbra to outward which confirms earlier reports \citep{Tziotziou2006, Nagashima2007, Socas-Navarro2009}. The power maps indicate that this decrease is set by the high-frequency contribution. Our analysis of spectral observations shows that the high frequency oscillations exist in the umbra of 5 min band. The observed power in the penumbral region shows strong peak in the 3 min band while the power is small at other frequencies. 

The penumbral oscillations are regarded as a tail of five minutes oscillations resulting further dependence of acoustic cutoff frequency on the magnetic field inclination \citep{Cally1994, DePontieu2004}. The oscillation power shows a strong relationship with the magnetic field strength and angle of inclination. Our analysis showed that the peak oscillation frequency not only depend upon the inclination but also on the field strength, although they behave differently in different frequency bands.

Observations in \Ha~and \Ca~bands demonstrated the oscillations with frequencies around 4.8\,mHz are dominant in the umbral and inner penumbral regions, together with the presence of RPWs in the band of 3\,mHz. Although the scale height between chromospheric \Ha~and \Ca~lines are closer to the photosphere, the relationship of frequency and magnetic fields may be affected. 

We can summarize the above results as follow: 
\begin{itemize}
	\item Ca\,{\sc ii} showed reverse ``C'' shaped bisectors at some locations of the sunspot.
	\item Bisector-measured Doppler velocity in \Ca~is affected by umbral emission features.
	\item Amplitude of RPWs decreases with distance from the umbra to outward and decay near the penumbral boundary.
	\item RPWs have much more blue than red in \Ha~and reverse in \Ca.
	\item Peak power frequency gradually increases from umbra to outward in both lines.
	\item In umbra and penumbra of sunspot, H$\alpha$ and \Ca~lines showed a phase difference of $\sim40^{\rm o}$ in the 3- and 5--minute oscillation bands.
	\item $v_{\rm rms}$ in the umbra is larger than penumbra of both lines. In outer umbra, \vrms~is larger in \Ha~than \vrms~in \Ca.
	\item The oscillation power shows different relation with the magnetic field strength and the inclination in different frequency bands.
\end{itemize}

We conclude that the main oscillations properties still suggest MHD waves propagating  upward along fanning field, as in earlier studies, but with interesting but yet unexplained response differences between the two  lines. The transformation of vertical oscillations into RPWs, at the boundary of umbra and penumbra, is an interesting feature which require further observations and theoretical studies to explain the wave phenomenon in the sunspot.

\begin{acks}
This work utilizes data from the Helioseismic and Magnetic Imager (HMI) on board Solar Dynamics Observatory (SDO). This work was support by the National Research Foundation of Korea (2011-0028102).
\end{acks}


\end{article}
\end{document}